\documentclass[12pt]{article}

\newcommand{\beq}{\begin{equation}}
\newcommand{\eeq}{\end{equation}}
\newcommand{\beqa}{\begin{eqnarray}}
\newcommand{\eeqa}{\end{eqnarray}}
\newcommand{\beqar}{\begin{eqnarray*}}
\newcommand{\eeqar}{\end{eqnarray*}}

\newcommand{\al}{\alpha}
\newcommand{\be}{\beta}

\def\spa          {\ \ \ }
\def\non          {\nonumber}
\def\ha           {\mbox{$\frac{1}{2}$}}

\def\spa          {\ \ \ }
\def\mand         {\spa\mbox{and}\spa}

\def\Tr           {\mbox{\rm Tr}\,}
\def\STr          {\mbox{\rm STr}\,}
\def\Str          {\mbox{\rm Str}\,}

\def\cd           {{\cdot}}
\def\ran          {\rangle}
\def\lan          {\langle}
\def\fsC    {C\!\!\!\!/\,}
\def\fsH    {H\!\!\!\!/\,}
\newcommand{\del}{\delta}

\newcommand{\eps}{\epsilon}
\newcommand{\ga}{\gamma}
\newcommand{\Ga}{\Gamma}

\newcommand{\inn}{\!\cdot\!}

\newcommand{\lam}{\lambda}

\newcommand{\ie}{{\it i.e.,}\ }
\newcommand{\labell}[1]{\label{#1}} 
\newcommand{\reef}[1]{(\ref{#1})}

\newcommand\cL{{\cal L}}
\newcommand\cD{{\cal D}}

\newcommand\cT{T}

\def\sst#1{{\scriptscriptstyle #1}}
\def\0{{\sst{(0)}}}
\def\1{{\sst{(1)}}}
\def\2{{\sst{(2)}}}
\def\3{{\sst{(3)}}}
\def\4{{\sst{(4)}}}
\def\5{{\sst{(5)}}}
\def\6{{\sst{(6)}}}
\def\7{{\sst{(7)}}}
\def\8{{\sst{(8)}}}

\begin{document}
\baselineskip 18pt%
\begin{titlepage}
\vspace*{1mm}%
\hfill
\vbox{

    \halign{#\hfil         \cr
           } 
      }  
\vspace*{8mm}

\center{ {\bf \Large   On Special Limit of Non-Supersymmetric Effective Actions of type II String theory

}}\vspace*{3mm} \centerline{{\Large {\bf  }}}
\begin{center}
{Ehsan Hatefi$^{a,b,c}$, Petr Vasko$^{a}$ }

\vspace*{0.6cm}{
Faculty of Physics, University of Warsaw, ul. Pasteura 5, 02-093 Warsaw, Poland$^{a}$\\
\quad\\

Scuola Normale Superiore and INFN, Piazza dei Cavalieri, 7, 56126 Pisa, Italy$^{b}$ \\
\quad\\

Mathematical Institute, Faculty of Mathematics,
Charles University, P-18675, CR$^{c,}$\footnote{ ehsan.hatefi@sns.it, vasko.petr@gmail.com, ehsanhatefi@gmail.com}  }

\end{center}
\begin{center}{\bf Abstract}\end{center}
\begin{quote}

In this paper we first address four point functions of string amplitudes in both type IIA and IIB string theories. Making use of non-BPS scattering amplitudes, we explore not only several Bianchi identities that hold in both transverse and world volume directions of brane, 
but also reveal various new couplings. These couplings can just be found by taking into account the mixed pull-back and Taylor couplings where their all order alpha-prime higher derivative corrections  have been derived as well.
 
For the first time, we also explore the complete form of a six point non-BPS amplitude, involving  three open string tachyons, a scalar field  and a Ramond-Ramond closed string  in both IIA, IIB.  In a special limit of the amplitude and using the proper expansion we obtain an infinite number of bulk singularities that are being constructed in the effective field theory. Finally  using new couplings  we  construct all the other massless  and tachyon singularities in type IIA, IIB string theories. All higher derivative corrections to these new couplings to  all orders in $\alpha'$ and new restricted Bianchi identities have also been gained.

 \end{quote}
\end{titlepage}

\section{Introduction}

Among several goals of theoretical physicists and in particular string theorists, we may point out to two common interests in uncovering more information about how the supersymmetry gets broken 
 as well as working out new couplings/ interactions  on time dependent backgrounds. If we try to deal with non-supersymmetric (unstable) branes, then one may be able to properly address some of the open questions and also might be able to deepen on many properties of various different string theories  \cite{Gutperle:2002ai,Lambert:2003zr,Sen:2004nf}. Since duality transformation is not promising in this context any more, one needs to be aware of  the fact that for non-BPS branes just  scattering amplitudes and Conformal Field Theory (CFT) methods \cite{Friedan:1985ge} would exactly determine all order $\alpha'$ corrections of the effective actions of string theory.

\vskip.1in

Making use of non-supersymmetric branes, the so called  Sakai-Sugimoto model \cite{Sakai:2004cn} as well as the symmetry breaking for holographic QCD models  have been  known \cite{Casero:2007ae}. Tachyons do play crucial role in  instability of the aforementioned systems so it would be important to consider tachyons and try to achieve their effective actions in both type IIA and IIB string theories and also explore their new couplings in the Effective Field Theory (EFT).

\vskip.1in

The leading order non-BPS effective actions including tachyonic modes  were proposed in \cite{Sen:1999md,Bergshoeff:2000dq}, where some of their properties such as  their decays and tachyon condensation have also been clarified in detail  \cite{Sen:2002in}. Following \cite{Sen:2004nf}, one reveals how to embed the presence of non-BPS branes in the effective actions. We studied D-brane anti-D-brane systems  \cite{Hatefi:2016yhb}. Recently the generalisation of effective actions of D-brane-anti-D-brane system to all orders in alpha-prime for both Chern-Simons and Dirac-Born-Infeld (DBI) effective actions was discovered \cite{Hatefi:2017ags}. Another example would be  related to tachyon condensation that has been investigated in  \cite{Sen:1998sm} in detail. For D-brane-anti-D-brane system, once  the distance between brane and anti-brane becomes smaller than the string length scale, two real tachyonic strings would appear. They are related to strings stretched from D-brane to anti-D-brane and vice versa. 

\vskip.1in
Here we would like to deal with N-coincident non-BPS branes and try to embed tachyonic modes and their corrections in EFT.  We take non-BPS scattering amplitude formalism as a theoretical framework or laboratory to discover their effective actions, including their all order  $\alpha'$ corrections in string theory in an efficient and  consistent way of matching string results  with EFT. To deal with the dynamics of unstable branes, we highlight  the recent work done by Polchinski and collaborators \cite{Michel:2014lva} where various explanations within the context of brane 's effective actions through EFT have been discussed. 
Not only brane production  \cite{Bergman:1998xv} but also inflation in string theory in the procedure of KKLT \cite{Dvali:1998pa} can also be mentioned. To observe a review of open strings and their features we point out to \cite{orientifold}.

\vskip.1in

In this paper we deal with a non-BPS four point function and explore some Bianchi identities as well as new EFT couplings that come from the mixed  Pull-back formalism and Taylor expansion
and then try to use the lower point functions to exactly build for the first time a non-BPS six point function. Having used the scattering amplitude methods, we would also fix some of the ambiguities  of the corrections in string theory and reveal  new string couplings in both type II string theories.\footnote{To work with some higher point functions and for their corrections  we just highlight \cite{Bjerrum-Bohr:2014qwa} and  \cite{Hatefi:2012zh,Hatefi:2012ve,Hatefi:2012rx} accordingly.}

One can try to relate some of the new couplings to AdS/CFT \cite{Hatefi:2012bp}. It is also worth making a remark on D-brane-Anti-D-brane system as they do affect not only in the problem of stability of KKLT model but also string compactifications \cite{Polchinski:2015bea} and in particular in the so called Large Volume Scenario. The relation between D-branes and Ramond-Ramond (RR) charges is well established 
 \cite{Polchinski:1995mt}, where  one could also take into account some brane's bound states
\cite{Witten:1995im}.  All the EFT methods of deriving  Wess-Zumino (WZ) and  DBI effective actions are given in \cite{Hatefi:2010ik,Myers:1999ps}.


 \vskip.2in

The paper is organised as follows. First we try to study a four point function including a closed string RR and a transverse scalar field and a real tachyon on the world volume of non-BPS branes, where an RR and two tachyons has been fully addressed in  \cite{Kennedy:1999nn}, then we build  all order $\alpha'$ higher derivative corrections to it and explore a pattern
from this calculation to reconstruct all singularity structures of higher point functions of non-BPS branes.

 \vskip.1in
Our notations for indices  are summarised by the following.

$\mu, \nu = 0, 1, ..., 9 $ represent the whole ten dimensional space-time, $a, b, c = 0, 1, ..., p$ show world volume indices and finally for transverse directions of the brane 
$ i, j = p + 1, ..., 9$ are taken accordingly.

\vskip.2in

We  establish a new coupling among RR, tachyon field living on the world volume of a non-BPS brane and one massless scalar field representing a transverse direction of the brane.

\vskip.3in

 This new mixed WZ -Taylor expansion is given by

\beqa
\frac{2i\beta'\mu_p'}{(p)!}(2\pi\alpha')^2 \int_{\Sigma_{p+1}} \epsilon^{a_{0}..a_{p}} C_{ia_{0}...a_{p-2}} D_{a_{p-1}}T D_{a_{p}} \phi^i \labell{highaa222}\eeqa

where $\mu_p'$ is RR charge of brane and $ \beta' $  is the WZ normalisation constant. 

Note that, the integration should be taken on $(p+1)$ world volume directions and in order to cover the whole world volume indices we extract the coupling and write it as \reef{highaa222}. We also explore its all order higher derivative corrections too.
\vskip.1in

Having set all lower point functions of non-BPS branes,  we would clarify more hidden symmetries in non-BPS context. Hence we make use of all the CFT techniques to a six point correlation of an RR, a scalar field and three tachyons. We first find out the entire correlators of $<V_{C^{-1}} V_{\phi^{-1}}V_{T^{0}} V_{T^{0}} V_{T^{0}}>$ in type IIA (IIB) and then we just illustrate the final result in different picture of scalar field, basically we explore $<V_{C^{-1}} V_{\phi^{0}}V_{T^{-1}} V_{T^{0}} V_{T^{0}}>$  and argue that using this particular case we would be able to precisely  obtain all bulk singularity structures that are not present in the other picture. Using selection rules \cite{Hatefi:2013yxa} for non-BPS amplitudes, EFT and in a particular soft limit, we discover the ultimate answer for the S-matrix. Having set all symmetries
  of the S-matrix, we explore the expansion of the S-matrix.  Using the soft limit  we generate not only all the infinite massless singularities but also an infinite number of u-channel bulk singularity structures can be precisely 
  reconstructed in an EFT and come to a perfect match between string amplitudes and EFT counterparts. Finally we use all the higher derivative corrections of two tachyon two scalar couplings 
to be able to produce an infinite number of scalar field singularities as well.  It is worth to emphasise since there is no coupling between two tachyon and a scalar field, the amplitude (as can be seen from the ultimate result of the S-matrix) has no 
singularity in $t,s,v$ channels at all. The DBI part of the effective action for non-BPS branes is 
\beqa
S_{DBI}&\sim&\int
d^{p+1}\sigma \STr\left(\frac{}{}V({ T^iT^i})\sqrt{1+\frac{1}{2}[T^i,T^j][T^j,T^i])}\right.\labell{nonab} \\
&&\qquad\qquad\left.
\times\sqrt{-\det(\eta_{ab}
+2\pi\alpha'F_{ab}+2\pi\alpha'D_a{ T^i}(Q^{-1})^{ij}D_b{ T^j})} \right)\,,\nonumber\eeqa  where $V({T^iT^i})=e^{-\pi{ T^iT^i}/2}$, and 
\beqa
Q^{ij}&=&I\delta^{ij}-i[T^i,T^j]\eeqa
 $i,j=1,2$, \ie $T^1=T\sigma_1$, $T^2=T\sigma_2$. The DBI part of the D-brane-anti-D-brane is given in
\cite{Hatefi:2017ags}. If we make kinetic terms symmetrized, find the traces and then use ordinary trace, the action  will get replaced to Sen's action \cite{Sen:2003tm}. However, in \cite{Garousi:2007fk,Hatefi:2012cp} by direct CFT computations and scattering amplitudes we have shown that Sen's effective action does not provide consistent result  with string amplitudes.  
The expansion of the S-matrices is consistent with Tachyon's 
potential  $V(|T|)=e^{\pi\alpha'm^2|T|^2}$ which comes from BSFT \cite{Kutasov:2000aq,Kraus:2000nj}. On the other hand WZ action is given by
 \beqa
S_{WZ}&=&\mu_p' \int_{\Sigma_{(p+1)}} C \wedge \Str e^{i2\pi\alpha'\cal F}\labell{WZ'}\eeqa 
To consider interactions with Tachyons, one can make contact with super connection of the non-commutative geometry ~\cite{quil,berl,Roepstorff:1998vh} where curvature is 
 \begin{displaymath}
i{\cal F} = \left(
\begin{array}{cc}
iF -\beta'^2 T^2 & \beta' DT \\
\beta' DT & iF -\beta'^2T^2 
\end{array}
\right) \ ,
\non\end{displaymath} 

One can find out different types of WZ  couplings from the above actions to generate consistent result between string amplitudes and the EFT, such as
\beqa
C\wedge \STr i{\cal F}&=& 2\beta' \mu_p'(2\pi\alpha') C_{p}\wedge DT\labell{exp2}\\
C\wedge \STr i{\cal F}\wedge i{\cal F}&=& \beta' \mu_p'(2\pi\alpha')^2 \bigg(C_{p-1}\wedge DT\wedge(DT)+C_{p-2}\wedge F \wedge DT\bigg)\nonumber
\eeqa

 \section{All order  $\alpha'$ corrections to $<V_{C^{-2}}  V_{T^{0}} V_{\phi^{0}}>$  }

 In order to actually address  entire form of a four point function of an RR, a real tachyon and a scalar field in both type IIA, IIB string theories, one must apply conformal field theory methods to the complete S-matrix elements and explore whether or not there are some bulk singularity structures  and also to notice  how one might be able to find out all order contact interactions. To achieve all the correlation functions, one needs to know vertex operators where their complete forms are shown by 
  \beqa
V_{T}^{(0)}(x) &=&  \alpha' ik_1\cd\psi(x) e^{\alpha' ik_1.\cd X(x)}\lam\otimes\sigma_1,
\nonumber\\
V_{T}^{(-1)}(x) &=&e^{-\phi(x)} e^{\alpha' ik_1\cd X(x)}\lam\otimes\sigma_2\nonumber\\
V_\phi^{(-1)}(x)&=&e^{-\phi(x)}\xi_{1i}\psi^i(x)e^{ \alpha'iq\inn X(x)}\lam\otimes \sigma_3 \nonumber\\
V_{\phi}^{(0)}(x) &=& \xi_{1i}(\partial^i X(x)+i\alpha'q.\psi\psi^i(x))e^{\alpha'iq.X(x)}\lam\otimes I\nonumber\\
V_{C}^{(-\frac{3}{2},-\frac{1}{2})}(z,\bar{z})&=&(P_{-}\fsC_{(n-1)}M_p)^{\alpha\beta}e^{-3\phi(z)/2}
S_{\al}(z)e^{i\frac{\alpha'}{2}p\cd X(z)}e^{-\phi(\bar{z})/2} S_{\be}(\bar{z})
e^{i\frac{\alpha'}{2}p\cd D \cd X(\bar{z})}\otimes\sigma_1,\nonumber\\
V_{C}^{(-\frac{1}{2},-\frac{1}{2})}(z,\bar{z})&=&(P_{-}\fsH_{(n)}M_p)^{\alpha\beta}e^{-\phi(z)/2}
S_{\al}(z)e^{i\frac{\alpha'}{2}p\cd X(z)}e^{-\phi(\bar{z})/2} S_{\be}(\bar{z})
e^{i\frac{\alpha'}{2}p\cd D \cd X(\bar{z})} \otimes\sigma_3\sigma_1\nonumber
\eeqa

Here $\lambda$ is the external Chan-Paton matrix for the
U(N) gauge group. The vertex operators of non-BPS D-branes should accompany internal degrees of freedom given the fact that if we send the tachyon to zero, one should recover the WZ action of BPS branes. For more information we recommend the section two of  \cite{Hatefi:2017oho} where $\sigma_i$ is Pauli Matrix.

This four point function at disk level can be computed if one takes into account the following  on-shell conditions 
\beqa
  q^2=p^2=0, \quad  k_{1}^2=1/4  , q.\xi_1=0,
\nonumber\eeqa
Projection operator and closed string RR 's field strength are defined by  \begin{displaymath}
P_{-} =\ha (1-\ga^{11}), \quad
\fsH_{(n)} = \frac{a
_n}{n!}H_{\mu_{1}\ldots\mu_{n}}\ga^{\mu_{1}}\ldots
\ga^{\mu_{n}},
\non\end{displaymath}

Spinor notation is given by $ (P_{-}\fsH_{(n)})^{\al\be} =
C^{\al\del}(P_{-}\fsH_{(n)})_{\del}{}^{\be}$ where C is charge conjugation matrix and for IIA (IIB) we pick up $n=2,4$,$a_n=i$  ($n=1,3,5$,$a_n=1$) accordingly. If we employ the doubling trick then one is able to just work out with holomorphic parts of the fields. Thus  we apply the the following change of variable to our field content 
\begin{displaymath}
\tilde{X}^{\mu}(\bar{z}) \rightarrow D^{\mu}_{\nu}X^{\nu}(\bar{z}) \ ,
\spa
\tilde{\psi}^{\mu}(\bar{z}) \rightarrow
D^{\mu}_{\nu}\psi^{\nu}(\bar{z}) \ ,
\spa
\tilde{\phi}(\bar{z}) \rightarrow \phi(\bar{z})\,, \mand
\tilde{S}_{\al}(\bar{z}) \rightarrow M_{\al}{}^{\be}{S}_{\be}(\bar{z})
 \ ,
\non\end{displaymath}

with 
\begin{displaymath}
D = \left( \begin{array}{cc}
-1_{9-p} & 0 \\
0 & 1_{p+1}
\end{array}
\right) \ ,\,\, \mand
M_p = \left\{\begin{array}{cc}\frac{\pm i}{(p+1)!}\ga^{i_{1}}\ga^{i_{2}}\ldots \ga^{i_{p+1}}
\eps_{i_{1}\ldots i_{p+1}}\,\,\,\,{\rm for\, p \,even}\\ \frac{\pm 1}{(p+1)!}\ga^{i_{1}}\ga^{i_{2}}\ldots \ga^{i_{p+1}}\ga_{11}
\eps_{i_{1}\ldots i_{p+1}} \,\,\,\,{\rm for\, p \,odd}\end{array}\right.
\non\end{displaymath}
\vskip .1in
Having carried the trick out, we would use the following propagators for all  $X^{\mu},\psi^\mu, \phi$ fields as follows 
\begin{eqnarray}
\lan X^{\mu}(z)X^{\nu}(w)\ran & = & -\frac{\alpha'}{2}\eta^{\mu\nu}\log(z-w) \ , \non \\
\lan \psi^{\mu}(z)\psi^{\nu}(w) \ran & = & -\frac{\alpha'}{2}\eta^{\mu\nu}(z-w)^{-1} \ ,\non \\
\lan\phi(z)\phi(w)\ran & = & -\log(z-w) \ .
\labell{prop2}\end{eqnarray}
Hence, our amplitude in the asymmetric picture of RR is found to be   \beqa
{\cal A}^{C^{-2}T^{0}\phi^{0}  }&=&\int dx_1 dx_2 dx_4 dx_5 (P_{-}\fsC_{(n-1)}M_p)^{\alpha\beta}(2i \alpha' k_{1a} \xi_{2i}) (x_{45})^{-3/4} (I_1+I_2)\nonumber\\&&\times
|x_{12}|^{\alpha'^2k_1.k_2}|x_{14}x_{15}|^{\frac{\alpha'^2}{2}k_1.p} |x_{24}x_{25}|^{ \frac{\alpha'^2}{2}  k_2.p}|x_{45}|^{\frac{\alpha'^2}{4}p.D.p}\nonumber\eeqa
with $x_4=z=x+iy,x_5=\bar z$ and 
\beqa
I_1&=&-ip^{i}\bigg(\frac{x_{45}}{x_{24}x_{25}}\bigg)
2^{-1/2}(x_{14}x_{15})^{-1/2}(x_{45})^{-3/4}(\gamma^{a} C^{-1})_{\alpha\beta}
\eeqa
To obtain the other correlation function including two spinors, a current and a fermion field $\bigg(I_2 = 2ik_{2b}<:S_{\al}(x_4): S_{\be}(x_5):\psi^{a}(x_1):\psi^{b}\psi^{i}(x_2):> \bigg)$
 we work out the so called Wick-like formula \cite{Liu:2001qa} to get to
\beqa
 I_2 &=& \bigg((\Gamma^{iba} C^{-1})_{\alpha\beta}-2\eta^{ab}(\gamma^{i} C^{-1})_{\alpha\beta} \frac{2Re[x_{14}x_{25}]}{x_{12}x_{45}}\bigg)
 \nonumber\\&&\times  2ik_{2b} 2^{-3/2}(x_{24}x_{25})^{-1}(x_{14}x_{15})^{-1/2}(x_{45})^{1/4}
  \nonumber\eeqa
One could precisely show that now the amplitude is  $SL(2,R)$ invariant and to remove the volume of conformal killing group we do gauge fixing as $(x_1,x_2,z,\bar z)=(x,-x,i,-i)$ with the Jacobian $J=-2i(1+x^2)$.
Setting the above gauge fixing, we come to know that  the second term of $I_2$ does not have any contribution to the final result of the amplitude due to the fact that integrand is odd while the moduli space is covered on the entire space-time or due to having symmetric interval. 
 $u = -\frac{\alpha'}{2}(k_1+k_2)^2$ is introduced and the amplitude is resulted by
 \beqa
{\cal A}^{C^{-2}T^{0}\phi^{0}  }&=&\int_{-\infty}^{\infty} dx (2x)^{-2u-1/2} 
(1+x^{2})^{-1/2 +2u} \bigg(p^i \Tr
(P_{-}\fsC_{(n-1)}M_p\gamma^{a})\nonumber\\&&+ik_{2b} \Tr
(P_{-}\fsC_{(n-1)}M_p\Gamma^{iba})\bigg) k_{1a} \xi_{2i}   \nonumber\eeqa

The ultimate result of amplitude is given by 
\beqa
{\cal A}^{C^{-2}T^0\phi^0} &=&
\bigg(p^i \Tr(P_{-}\fsC_{(n-1)}M_p\gamma^{a})+ik_{2b} \Tr
(P_{-}\fsC_{(n-1)}M_p\Gamma^{iba})\bigg) k_{1a} \xi_{2i} \nonumber\\&&\times (\pi\beta'\mu_p')2\sqrt{\pi} \frac{\Ga[-u+1/4]}{\Ga[3/4-u]} \labell{amp383}\ .
\eeqa
  $ \mu_p' $ is RR charge of brane. All the traces are non zero for $p+1= n$ case and can be calculated as
\beqa
\Tr\bigg(\fsC_{(n-1)}M_p (k_1.\ga)\bigg)&=&\pm\frac{32}{p!}\eps^{a_{0}\cdots a_{p-1}a}C_{a_{0}\cdots a_{p-1}}k_{1a} \nonumber\\
\Tr\bigg(\fsC_{(n-1)}M_p (\xi.\ga)(k_2.\ga)(k_1.\ga)\bigg)&=&\pm\frac{32}{p!}\eps^{a_{0}\cdots a_{p-2}ba}C_{a_{0}\cdots a_{p-2}}k_{1a} k_{2b} \xi_{1i}
\nonumber\eeqa
 The correct expansion of the amplitude can be found by dealing with either massless or tachyon poles of the amplitude. From a three point function including an RR and a real tachyon and using its momentum conservation along the world volume of brane $k^2=p^ap_a=\frac{1}{4} $\cite{Hatefi:2012wj}, one realises that this constraint holds for $CT\phi$ amplitude and indeed the proper momentum expansion can be read off as follows
 \beqa
u=-p^ap_a\rightarrow \frac{-1}{4},\quad  \sqrt{\pi}\frac{\Ga[-u+1/4]}{\Ga[3/4-u]}
=\pi \sum_{n=-1}^{\infty}c_n(u+1/4)^{n+1}
\ .\labell{taylor61}\nonumber
\eeqa
where the 1st three coefficients are
\beqa
c_{-1}&=&1,c_0=2ln(2),c_1=\frac{1}{6}(\pi^2+12ln(2)^2).\nonumber\eeqa

The first term in \reef{amp383} can be produced by using the following Chern-Simons coupling where the scalar field has been taken from the Taylor expansion
\beqa
S_1&=&\frac{2i\beta'\mu_p'}{p!}(2\pi\alpha')^2 \int_{\Sigma_{p+1}} \partial_i C_{p}\wedge DT\phi^i \labell{highaa}\eeqa
 
 Note that the second term of \reef{amp383} can just be produced if one introduces a new coupling  where this time a scalar field comes from pull-back of brane and covariant derivative of tachyon is appeared to cover the entire $(p+1)$  world volume direction. Hence the second term of \reef{amp383} can be regenerated by the following new mixed WZ and pull-back coupling

 \beqa
S_2&=&\frac{2i\beta'\mu_p'}{(p)!}(2\pi\alpha')^2 \int_{\Sigma_{p+1}} \epsilon^{a_{0}..a_{p}} C_{ia_{0}...a_{p-2}} D_{a_{p-1}}T D_{a_{p}} \phi^i \labell{highaa22}\eeqa


As we have seen, the expansion of the amplitude has an infinite contact interaction and all those contact interaction terms related to the first term of \reef{amp383} can be produced in the EFT by applying all infinite higher derivative corrections to the WZ effective actions of a real tachyon, a scalar field and a $C_{p}$ RR closed string  \reef{highaa}
\beqa
\frac{2i\beta'\mu_p'}{p!}(2\pi\alpha')^2 \int_{\Sigma_{p+1}} \partial_i C_{p} \wedge \Tr \bigg(\sum_{n=-1}^{\infty}c_n(\alpha')^{n+1}  D_{a_1}\cdots D_{a_{n+1}}DT D^{a_1}...D^{a_{n+1}}\phi^i \bigg)
 \labell{highaa33}\eeqa
 Likewise all the contact interactions related to the second term \reef{amp383} can be constructed if one applies 
  the same prescription to all higher derivative corrections to $S_2$ action as follows

 \beqa
\frac{2i\beta'\mu_p'}{p!}(2\pi\alpha')^2 \int_{\Sigma_{p+1}}       \epsilon^{a_{0}..a_{p}}  C_{ia_{0}...a_{p-2}} \Tr\bigg(\sum_{n=-1}^{\infty}c_n(\alpha')^{n+1}  D_{a_1}\cdots D_{a_{n+1}}
 D_{a_{p-1}}T D^{a_1}...D^{a_{n+1}} D_{a_{p}}\phi^i\bigg) \nonumber\eeqa

It is also interesting to revisit the amplitude in the other pictures. The final result of the amplitude can be derived as  
\beqa
{\cal A}^{C^{-1}T^0\phi^{-1}} &=&2 \Tr(P_{-}\fsH_{(n)}M_p\Gamma^{ia}) k_{1a} \xi_{2i} (\pi\beta'\mu_p') \sqrt{\pi} \frac{\Ga[-u+1/4]}{\Ga[3/4-u]} \labell{amp3844}\ .
\eeqa
The trace that includes $\gamma^{11}$ factor, has the special property  so that all results are being held for the following relations as well 
\beqa
  p>3 , H_n=*H_{10-n} , n\geq 5.
  \nonumber\eeqa
Now if we apply momentum conservation $(k_1+k_2+p)^a=0$ to the above amplitude then we realise that the amplitude \reef{amp3844} can just produce the 1st term of \reef{amp383},  more importantly one finds that a Bianchi identity holds for the world volume of branes  in the presence of RR's field strength as
\beqa
  p_a H_{a_0...a_{p-1}}\epsilon^{a_0...a_{p-1}a}=0 \label{jj33}\eeqa 
Finally the result of the amplitude for  ${\cal A}^{T^{-1}\phi^0 C^{-1}}$ is derived to be
  \beqa
 \bigg(k_{2b} \Tr(P_{-}\fsH_{(n)}M_p\Gamma^{ib})-p^i \Tr(P_{-}\fsH_{(n)}M_p)\bigg)\xi_{2i} (2\pi\beta'\mu_p') \sqrt{\pi} \frac{\Ga[-u+1/4]}{\Ga[3/4-u]}  \nonumber \eeqa
 
Finding the above result and keeping in mind momentum conservation, one understands that  to get the consistent result with both string theory and effective field theory parts, the restricted world volume Bianchi identity \reef{jj33} has  to be modified by a new Bianchi identity which will be valid for both world volume and transverse directions of the branes as follows
 \beqa
  p^i \epsilon^{a_0...a_{p}}H_{a_0...a_{p}}+p^a \epsilon^{a_0...a_{p-1}a}H^{i}_{a_0...a_{p-1}}=0
  \label{jj66}\eeqa

\section{$<V_{C^{-1}}V_{\phi^{-1}}V_{T^{0}}V_{T^{0}}V_{T^{0}}>$ amplitude}
 In this section we would like to deal with a non-BPS six point function including an RR, a transverse scalar field and three real tachyons to be able to find not only the proper expansion of the amplitude but also reveal all the bulk singularity structures as well as various restricted Bianchi  identities. Given the exact symmetries of string theory amplitudes, tachyon expansion and the particular soft limit,  in the following we show that one is able to predict some of the singularity structures of $<V_{C^{-1}(z,\bar z)}V_{\phi^{-1} (x_1)}V_{T^{0}(x_2)}V_{T^{0}(x_3)}V_{T^{0}(x_4)}>$ amplitude. We then work out with $<V_{C^{-1}(z,\bar z)}V_{\phi^{0} (x_1)}V_{T^{-1}(x_2)}V_{T^{0}(x_3)}V_{T^{0}(x_4)}>$ and determine all the singularities including the bulk singularities that carry momentum of RR in the bulk directions. One needs to provide the correlation function between two spinors and four fermion fields at different locations where just one of them moves along transverse direction of brane so $I_1^{cbai}=<:S_{\al}(x_z): S_{\be}(x_{\bar z}):\psi^{i}(x_1):\psi^{a}(x_2):\psi^{b}(x_3):\psi^{c}(x_4):>$ is found to be
 \beqa
I_1^{cbai}&=&
\bigg\{(\Gamma^{cbai}C^{-1})_{{\alpha\beta}}-\alpha' \eta^{ab}(\Gamma^{ci}C^{-1})_{\alpha\beta}\frac{Re[x_{25}x_{36}]}{x_{23}x_{56}}+\alpha' \eta^{ac}(\Gamma^{bi}C^{-1})_{\alpha\beta}\frac{Re[x_{25}x_{46}]}{x_{24}x_{56}}
\nonumber\\&& -\alpha' \eta^{bc}(\Gamma^{ai}C^{-1})_{\alpha\beta}\frac{Re[x_{35}x_{46}]}{x_{34}x_{56}}\bigg)
\bigg\}2^{-2}x_{45}^{3/4}(x_{15}x_{16}x_{25}x_{26}x_{35}x_{36}x_{45}x_{46})^{-1/2}
\nonumber\eeqa
 
Note that here $x_5=z=x+iy, x_6=\bar z$.
 All the techniques have already been explained, fixing the position of open strings at $x_1=0, 0\leq x_2\leq 1 , x_3=1, x_4=\infty$ and using 6 independent Mandelstam variables as $s=-(\frac{1}{4}+2k_1.k_3), t=-(\frac{1}{4}+2k_1.k_2), v=-(\frac{1}{4}+2k_1.k_4), u=-(\frac{1}{2}+2k_2.k_3), r=-(\frac{1}{2}+2k_2.k_4), w=-(\frac{1}{2}+2k_3.k_4)$ the final form of the amplitude is written by

\beqa 
{\cal A}&=&4i\xi_{1i} (P_{-}\fsH_{(n)}M_p)^{\alpha\beta}\int_{0}^{1} dx_2 x_2^{-2t-1/2} (1-x_2)^{-2u-1}\int dz \int d\bar z |1-z|^{2s+2u+2w+1/2} |z|^{2t+2s+2v-1/2}
\nonumber\\&&\times k_{2a}k_{3b}k_{4c}(z - \bar{z})^{-2(t+s+u+v+r+w)-5/2}  |x_2-z|^{2t+2u+2r+1/2}
\bigg[(\Gamma^{cbai}C^{-1})_{\alpha\beta}+(z - \bar{z})^{-1}\nonumber\\&&\times\bigg( 
2\eta^{ab}(\Gamma^{ci}C^{-1})_{\alpha\beta}(1-x_2)^{-1}(x_2-xx_2-x+|z|^{2})
-2\eta^{ac}(\Gamma^{bi}C^{-1})_{\alpha\beta}(x_2-x)
\nonumber\\&&+2\eta^{bc}(\Gamma^{ai}C^{-1})_{\alpha\beta}(1-x)\bigg)\bigg] \label{eerr33}\eeqa
The amplitude makes sense for $p=n+1,p+1=n$ cases. The algebraic form of the above integrals 
can be derived in a soft limit $4k_2.p \rightarrow 1$. Using this limit and appendix B of \cite{Hatefi:2012wj} and \cite{Fotopoulos:2001pt} one arrives at closed form for the integrals.  For the simplicity, we just write down
the ultimate result of the amplitude for $p=n+1$ case as  \beqa
{\cal A}_{1}^{C\phi TTT}&=&4i\xi_{1i} \pi
k_{2a}k_{3b}k_{4c}
\Tr(P_{-}\fsC_{(n-1)}M_p\Gamma^{cbai})M_1 M_2
\labell{bb44331}\eeqa
    
 where $M_1,M_2$  are 
 \beqa 
M_1&=&(2)^{-2(t+s+u+v+r+w)-5/2}{\frac{\Gamma(-2t+\frac{1}{2})\Gamma(-2u)}
{\Gamma(-2t-2u+\frac{1}{2})}}\nonumber\eeqa
\beqa
M_2={\frac{\Gamma(-u-r-w-\frac{1}{2})\Gamma(-t-v-r)\Gamma(-s+r+\frac{1}{4})\Gamma(-t-s-u-v-r-w-\frac{3}{4})}{\Gamma(-u-s-w-\frac{1}{4})\Gamma(-t-s-v+\frac{1}{4})\Gamma(-u-w-t-v-2r-\frac{1}{2})}}\nonumber\eeqa

The other part of the amplitude holds for $C_{p}$ case and one reveals its final form as follows
\beqa
{\cal A}_{2}&=& M_1\pi \frac{32}{(p+1)!}\epsilon^{a_0...a_{p-1}a} H^{i}_{a_0...a_{p-1}}\xi_{1i} i\bigg\{-k_{2a}(w+\frac{1}{2})(-r-t-v-\frac{1}{2})M_3\nonumber\\&&+\frac{1}{4(-2t-2u+\frac{1}{2})}k_{3a}(r+\frac{1}{2})M_3
\bigg((-1+r(-2+8t-8u)-2v+2t(1+4t+4v) \nonumber\\&&-8u(1+u+w))\bigg)
+\frac{1}{16}k_{4a}M_4\bigg(4s(-1+4t)+4(5+4r)u+8r+3+20t\nonumber\\&&+4w+16(t+u)(u+w)\bigg) \bigg\},\label{amp1aa}\eeqa
where $M_3,M_4$ are written in terms of ratio of the Gamma functions
\beqa
M_3&=&{\frac{
\Gamma(r-s+\frac{3}{4})\Gamma(-t-v-r-\frac{1}{2})\Gamma(-u-r-w-1)\Gamma(-t-s-u-v-r-w-\frac{5}{4})}
{\Gamma(-t-s-v+\frac{1}{4})\Gamma(-u-s-w-\frac{1}{4})\Gamma(-t-u-v-w-2r-\frac{1}{2})}}\nonumber\\
M_4&=&{\frac{
\Gamma(r-s-\frac{1}{4})\Gamma(-t-v-r+\frac{1}{2})\Gamma(-u-r-w-1)\Gamma(-t-s-u-v-r-w-\frac{5}{4})}
{\Gamma(-t-s-v+\frac{1}{4})\Gamma(-u-s-w-\frac{1}{4})\Gamma(-t-u-v-w-2r-\frac{1}{2})}}\nonumber
\eeqa
Let us deal with bulk singularities.
\section{$<V_{C^{-1}(z,\bar z)}V_{\phi^{0} (x_1)}V_{T^{-1}(x_2)}V_{T^{0}(x_3)}V_{T^{0}(x_4)}>$ amplitude}

In this section we would like to  produce all the massless bulk singularity structures that carry momentum of RR in the transverse directions. To do so, we deal with the following  $<V_{C^{-1}(z,\bar z)}V_{\phi^{0} (x_1)}V_{T^{-1}(x_2)}V_{T^{0}(x_3)}V_{T^{0}(x_4)}>$ amplitude. All the correlation functions can be computed. To shorten the paper we use the same gauge fixing as in the last section. Thus the final form of the amplitude is found     
\beqa 
{\cal A}&\sim&4i\xi_{1i} (P_{-}\fsH_{(n)}M_p)^{\alpha\beta}\int_{0}^{1} dx_2 x_2^{-2t-1/2} (1-x_2)^{-2u-1}\int dz \int d\bar z |1-z|^{2s+2u+2w+1/2} |z|^{2t+2s+2v-1/2}
\nonumber\\&&\times k_{3b}k_{4c}(z - \bar{z})^{-2(t+s+u+v+r+w)-5/2}  |x_2-z|^{2t+2u+2r+1/2}
\bigg[p^i \bigg(2\eta^{bc}(C^{-1})_{\alpha\beta}\frac{1-x}{(z-\bar z)}\nonumber\\&&+(\Gamma^{cb}C^{-1})_{\alpha\beta}\bigg)
+k_{1a}\bigg((\Gamma^{cbia}C^{-1})_{\alpha\beta}+(z - \bar{z})^{-1}\bigg( x  l_1+ l_2 |z|^{2}+l_3\bigg)\bigg)\bigg] \label{ee66}\eeqa
where
  \beqa 
l_1&=&2\eta^{ab}(\Gamma^{ci}C^{-1})_{\alpha\beta}-2\eta^{ac}(\Gamma^{bi}C^{-1})_{\alpha\beta}-2\eta^{bc}(\Gamma^{ia}C^{-1})_{\alpha\beta}\nonumber\\
l_2&=&-2\eta^{ab}(\Gamma^{ci}C^{-1})_{\alpha\beta}\nonumber\\
l_3&=&2\eta^{bc}(\Gamma^{ia}C^{-1})_{\alpha\beta}
\nonumber\eeqa
The amplitude makes sense for $p=n+1,p+1=n$ cases. Using soft limit $4k_2.p \rightarrow 1$, one finds the amplitude for $p=n+1$ case in below
\beqa
{\cal A}_{1}^{C^{-1}\phi^{0} T^{-1}T^0T^0}\sim \frac{64i\xi_{1i} \pi M_1 M_2 k_{3b}k_{4c}}{(p-1)!}\bigg(k_{1a}\epsilon^{a_0...a_{p-3}acb} H^{i}_{a_0...a_{p-3}}+p^i \epsilon^{a_0...a_{p-2}cb} H_{a_0...a_{p-2}}\bigg)\labell{bb172}\eeqa
    
 \vskip.2in

The second part of the amplitude holds for $C_{p}$ case and one finds it as \beqa
{\cal A}_{2}&=& M_1\pi \frac{32}{(p)!}\xi_{1i} i\bigg\{ -p^i \epsilon^{a_0...a_{p}} H_{a_0...a_{p}}(w+\frac{1}{2})(-r-t-v-\frac{1}{2})M_3\nonumber\\&&
+ \epsilon^{a_0...a_{p-1}d} H^{i}_{a_0...a_{p-1}} M_4\bigg(k_{4d}(s+\frac{1}{4})(-u-r-w-1) 
-k_{1d}(w+\frac{1}{2})(r-s-\frac{1}{4})\nonumber\\&&
+k_{3d}\frac{(r-s-\frac{1}{4}) }{(-r-t-v-\frac{1}{2})}(v+\frac{1}{4})(-u-r-w-1) \bigg)\bigg\},\label{533}
\eeqa
Now let us deal with all singularities and start to produce them.

\section{Bulk Singularity Structures}

To obtain all the singularity structures including the ones that carry momentum of the closed string RR in the bulk direction, one needs to find first the expansion of the amplitude. The following remarks need to be taken into consideration.  
Having applied momentum conservation we get  $ s+t+u+v+r+w=-p^a p_a-\frac{3}{2}$. Using the constraint $p^ap^a\rightarrow \frac{1}{4}$ for non-BPS branes, taking the symmetries of our amplitude (given the EFT and the fact that it should be symmetric under exchanging $s,t,v$), one gains the expansion for the amplitude \reef{533} as follows
 \beqa
(s,t,v\rightarrow -\frac{1}{4}), (u,r\rightarrow 0), w\rightarrow -1\label{expas1}
\eeqa

The expansion of $(M_1 M_2)$  for this particular soft limit  and around \reef{expas1} can be read as 
\beqa
-\frac{\pi^{3/2}}{u}-\frac{\pi^{7/2}}{6u}\bigg((t+s+r+w)^2+2t(v-w)-2s(t+w)+2r(r+v)+v^2+...\bigg)
\labell{esiqq}\eeqa

Given the above expansion, standard EFT methods that propose to us to have an infinite u-channel massless gauge field poles and symmetries \cite{Schwarz:2013wra}, one understands $k_i.k_j\rightarrow 0$ for massless strings 
 and $p^ap^a\rightarrow \frac{1}{4}$ for non-BPS branes. There is no coupling between two tachyon and a scalar field and given the selection rules  for non-BPS branes  \cite{Hatefi:2013yxa}, the fact that the kinetic term of tachyon is fixed in DBI action, one clarifies that there is no double pole for $p=n+1$ case. This point can also be observed from the expansion of the amplitude, hence we have an infinite number of u-channel simple poles. For  $p=n+1$ case, from EFT and the above expansion, one notices that the S-matrix has   an infinite number of u-channel gauge field poles. The first u-channel pole in string theory can be written down as 
 \beqa
\frac{64i \pi^{3/2}}{(p-1)!u}\epsilon^{a_0\cdots a_{p-3}acb}H^{i}_{a_0\cdots a_{p-3}} k_{1a}k_{3b}k_{4c}\xi_{1i}   \Tr(\lambda_1\lambda_2\lambda_3\lambda_4)\label{765}
\eeqa
This simple u-channel pole can be reconstructed in an EFT by the following sub amplitude
 \beqa
 && V^{\alpha}_{a}(T_2,T_3,A) G^{\alpha\beta}_{ab}(A)V^{\beta}_{b}(C_{p-2},A,\phi_1,T_4)\labell{vvxz22}\\
V^{\alpha}_{a}(T_2,T_3,A)&=&i T_p (2\pi\alpha') (k_2-k_3)_{a} \Tr(\lambda_2\lambda_3\lambda^{\alpha})\nonumber\\
G^{ab}(A) &=&\frac{i\delta^{ab} \delta^{\alpha\beta}}{(2\pi\alpha')^2 T_p u}\nonumber\\
V^{\beta}_{b}(C_{p-2},A,\phi_1,T_4)&=&i\mu'_p\beta' (2\pi\alpha')^3\frac{1}{(p-1)!}\epsilon^{a_0\cdots a_{p-1}b}H^{i}_{a_0\cdots a_{p-3}} k_{4a_{p-2}}k_{1a_{p-1}}\xi_{1i}  \Tr(\lambda_4\lambda_1\lambda^{\beta})\nonumber
\eeqa
Here $\alpha,\beta$ are gauge group indices. Now if we use the above vertices in an EFT and make use of momentum conservation $(k_1+k_2+k_3+k_4+p)^a=0$, Bianchi identity $p_{a_{p-2}} H_{a_0\cdots a_{p-3}}=0$, the fact that the amplitude is symmetric under $k_{1a_{p-1}}k_{1a_{p-2}}$ and also due to antisymmetric property of $\epsilon$, we realise that the term $k_{1a_{p-1}}k_{1a_{p-2}}$ does not have any effect in EFT part of the amplitude. If we multiply \reef{765} by $\frac{1}{2}\mu'_p\beta' \pi^{1/2}$ and compare it with above EFT amplitude we then explore that the first simple u-channel gauge field pole is exactly generated.

$V^{a}(C_{p-2},A,\phi_1,T_4)$  was derived from some part of the mixed WZ coupling and Taylor expansion 
 \beqa
 \beta'\mu'_p (2\pi\alpha')^{3}\int_{\Sigma_{p+1}}\Tr (\partial_i C_{p-2}\wedge F\wedge DT \phi^i)\label{bb123}\eeqa  
 
 However, as can be seen from the expansion, the amplitude in \reef{bb172} has an infinite number of bulk singularity structures concretely, where the first bulk pole in string theory reads as

 \beqa
 \frac{32i  \beta'\mu'_p \pi^2\xi_{1i}  k_{3b}k_{4c}}{(p-1)!u} p^i \epsilon^{a_0...a_{p-2}cb} H_{a_0...a_{p-2}}\labell{bni1}\eeqa

 If one wants to extract the couplings from \reef{bb123}, one takes integration by part to arrive at two different contributions   
  \beqa
 -\beta'\mu'_p (2\pi\alpha')^{3}\int_{\Sigma_{p+1}} \epsilon^{a_0...a_p}\bigg( d_{a_{p-2}} \partial_i C_{a_0...a_{p-3}} A_{a_{p-1}} d_{a_p} T \phi^i- \partial_i C_{a_0...a_{p-3}} A_{a_{p-1}} d_{a_p} T d_{a_{p-2}}\phi^i\bigg)\label{ll1}\eeqa  
 where to derive the first simple u-channel gauge pole given in \reef{765}, we have already used the contribution from the second term of \reef{ll1}. Now if we use the following relation 
 \beqa
 (p-2)\partial_i C_{a_0...a_{p-3}}= H^{i}_{a_0...a_{p-3}}-\partial_{[a_{p-3}} C_{a_0...a_{p-4}]i} \label{661q}
 \eeqa
and plug it into the first term of \reef{ll1}, then we would be able to produce the bulk part of the vertex  $V^{b}(C_{p-2},A,\phi_1,T_4)$. Eventually by replacing it into the same EFT sub amplitude \reef{vvxz22} we are able to precisely produce the first bulk singularity u-channel pole \reef{bni1} which carries $p.\xi$ term as well. As one notices from the expansion of the amplitude,  we have an infinite number of u-channel poles and to generate them in an EFT the following remarks are in order.

 The vertex of $V^{\alpha}_{a}(T_2,T_3,A)$ comes from the kinetic term of tachyons in DBI action which is fixed and has no correction. The propagator is also fixed, as it comes from the kinetic term of gauge fields that has been fixed for this case too. Therefore to be able to reconstruct all infinite u-channel poles one must directly apply infinite higher derivative corrections to the mixed WZ coupling \reef{bb123}  as follows
  \beqa
 \beta'\mu'_p (2\pi\alpha')^{3} \sum_{n=-1}^{\infty}b_n \int_{\Sigma_{p+1}}\Tr \bigg(\partial_i C_{p-2}\wedge  D^{b_1}\cdots D^{b_{n}} F \wedge D_{b_1}\cdots D_{b_{n}}\bigg[ DT \phi^i\bigg]\bigg).\label{bb1298}\eeqa 
 
Keeping fixed the 
$ V^{\alpha}_{a}(T_2,T_3,A) $ and propagator, extracting the modified all order vertex  $V^{b}(C_{p-2},A,\phi_1,T_4)$ from \reef{bb1298} and replacing it in the EFT amplitude  \reef{vvxz22}, then one is able to show that all infinite bulk singularity structures are precisely produced. This clearly confirms that the expansion is consistent with EFT amplitude as well.


The amplitude has an infinite number of massless scalar poles in $(t'+v'+r)$ \footnote{$t'=t+\frac{1}{4},v'=v+\frac{1}{4}$} channel that correspond to the extensions of higher derivative corrections of two tachyon-two scalar field couplings. These corrections originate from the second part of the amplitude in \reef{533}. They are reconstructed by the following EFT  prescription  \beqa
{\cal A}&=&V^{\alpha}_{i}(C_{p},T_3,\phi)G^{\alpha\beta}_{ij}(\phi)V^{\beta}_{j}(\phi,T_2,\phi_1,T_4)\nonumber\\
G^{\alpha\beta}_{ij}(\phi) &=&\frac{i\delta_{\alpha\beta}\delta_{ij}}{(2\pi\alpha')^2 T_p (t'+v'+r)}\nonumber\\
V^{\alpha}_{i}(C_{p},T_3,\phi)&=&\mu'_p\beta' (2\pi\alpha')^2\frac{1}{(p)!}\epsilon^{a_0\cdots a_{p}}H^{i}_{a_0\cdots a_{p-1}} k_{3a_{p}}\Tr(\lambda_3\Lambda^{\alpha})\labell{4482vv}\eeqa

To generate all scalar poles at first order one needs to employ the following Lagrangian
\beqa
{\cal{L}}(\phi,\phi,T,T)&=& -2T_p(\pi\alpha')^3{\rm
STr} \bigg(
m^2\cT^2(D_a\phi^iD^a\phi_i)+\frac{\alpha'}{2}D^{\alpha}\cT D_{\alpha}\cT D_a\phi^iD^a\phi_i \nonumber\\&&-
\alpha' D^{b}\cT D^{a}\cT D_a\phi^iD_b\phi_i \bigg)\labell{dbicoupling} \eeqa
while  to produce all the other poles, one needs to know higher derivative corrections to the two tachyon-two scalar field couplings  to all orders
\beqa
\cL_{}&=&-2T_p(\pi\alpha')(\alpha')^{2+n+m}\sum_{n,m=0}^{\infty}(\cL_{1}^{nm}+\cL_{2}^{nm}+\cL_{3}^{nm}+\cL^{nm}_{4}),\labell{lagrango}\eeqa
where
\beqa
\cL_1^{nm}&=&m^2
\Tr\left(\frac{}{}a_{n,m}[\cD_{nm}(\cT^2 D_a\phi^iD^a\phi_i)+ \cD_{nm}(D_a\phi^iD^a\phi_i\cT^2)]\right.\nonumber\\
&&\left.+\frac{}{}b_{n,m}[\cD'_{nm}(\cT D_a\phi^i\cT D^a\phi_i)+\cD'_{nm}( D_a\phi^i\cT D^a\phi_i\cT)]+h.c.\right),\nonumber\\
\cL_2^{nm}&=&\Tr\left(\frac{}{}a_{n,m}[\cD_{nm}(D^{\alpha}\cT D_{\alpha}\cT D_a\phi^iD^a\phi_i)+\cD_{nm}( D_a\phi^iD^a\phi_i D^{\alpha}\cT D_{\alpha}\cT)]\right.\nonumber\\
&&\left.+\frac{}{}b_{n,m}[\cD'_{nm}(D^{\alpha} \cT D_a\phi^i D_{\alpha}\cT D^a\phi_i)+\cD'_{nm}( D_a\phi^i D_{\alpha}\cT D^a\phi_i D^{\alpha} \cT)]+h.c.\right),\nonumber\\
\cL_3^{nm}&=&-\Tr
\left(\frac{}{}a_{n,m}[\cD_{nm}(D^{\beta}\cT D_{\mu}\cT D^\mu\phi^iD_\beta\phi_i)+\cD_{nm}( D^\mu\phi^iD_\beta\phi_iD^{\beta}\cT D_{\mu}\cT)]\right.\nonumber\\
&&\left.+\frac{}{}b_{n,m}[\cD'_{nm}(D^{\beta}\cT D^\mu\phi^iD_{\mu}\cT D_\beta\phi_i)+\cD'_{nm}(D^\mu\phi^i D_{\mu}\cT  D_\beta\phi_i  D^{\beta}\cT)]+h.c.\right),\nonumber\\
\cL_4^{nm}&=&-\Tr\left(\frac{}{}a_{n,m}[\cD_{nm}(D^{\beta}\cT D^{\mu}\cT D_\beta\phi^iD_\mu\phi_i)
+\cD_{nm}( D^\beta\phi^iD^\mu\phi_iD_{\beta}\cT D_{\mu}\cT)]\right.\nonumber\\
&&\left.+\frac{}{}b_{n,m}[\cD'_{nm}(D^{\beta}\cT D_\beta\phi^iD^{\mu}\cT D_\mu\phi_i)+\cD'_{nm}( D_\beta\phi^i D_{\mu}\cT  D^\mu\phi_i D^{\beta}\cT)]+h.c.
\right).\label{hdts}
\eeqa
The definitions for $D_{nm}$ and $D'_{nm}$ higher derivative operators are
\beqa
D_{nm}(EFGH)&\equiv&D_{b_1}\cdots D_{b_m}D_{a_1}\cdots D_{a_n}E  F D^{a_1}\cdots D^{a_n}GD^{b_1}\cdots D^{b_m}H\nonumber\\
D'_{nm}(EFGH)&\equiv&D_{b_1}\cdots D_{b_m}D_{a_1}\cdots D_{a_n}E   D^{a_1}\cdots D^{a_n}F G D^{b_1}\cdots D^{b_m}H\nonumber\eeqa
 The all order extended vertex $V^{\beta}_{j}(\phi,T_2,\phi_1,T_4)$ is derived from \reef{hdts} and in momentum space takes the form  
\beqa 
V_{\beta}^{j}(\phi,\phi_1, T_2,T_4)&=& \frac{1}{2}v't'\xi_{1}^j (-2i
T_p\pi)(\alpha')^{n+m+3}(a_{n,m}+b_{n,m}) \bigg(\frac{}{}(k_2\inn
k_1)^n(k_1\inn k_4)^m+(k_2\inn k_1)^m(k_4\inn k_1)^n
\nonumber\\&&+(k_2\inn k_1)^n(k\inn k_2)^m+(k_2\inn k_1)^m (k\inn
k_2)^n  +(k_1\inn k_4)^m(k\inn k_4)^n+(k_1\inn k_4)^n(k\inn
k_4)^m\nonumber\\&& +(k\inn k_2)^m(k\inn k_4)^n+(k\inn
k_2)^n(k\inn k_4)^m \bigg) \Tr(\lam_4\lam_1\lam_2\lambda_{\beta})\labell{verpptt}\eeqa
 where $k$ is the momentum of the off-shell scalar field.  Substituting the above vertex in the EFT amplitude \reef{4482vv}, we produce all infinite scalar poles as 

\beqa
&&8i\mu'_p \beta' \frac{\eps^{a_{0}\cdots a_{p}}\xi_{1i} H^{i}_{a_0\cdots
a_{p-1}}k_{3a_{p}}}{p!(v'+t'+r)}\Tr(\lam_1\lam_2\lam_3\lam_4)
\sum_{n,m=0}^{\infty}(a_{n,m}+b_{n,m})[v'^{m}t'^{n}+v'^{n}t'^{m}]
  v't' \label{amphigh8}\eeqa
Eventually, the S-matrix suggests to us that the string amplitude has just a double pole for $p+1=n$ case. It emerges from the following Feynman diagram in EFT \footnote{ we suppress all gauge indices.}

\beqa
 V(C_{p},\phi_1,T)G(T)V_{a}(T,T_4,A)G_{ab}(A)V_b(A,T_2,T_3)\labell{amp777xz}\eeqa
 with the derived vertices taken from the  lower order effective actions \beqa
V(C_{p},\phi_1,T)&=&\beta'\mu'_p(2\pi\alpha')^2\frac{1}{(p+1)!} p^i \epsilon^{a_0...a_{p}} H_{a_0...a_{p}}\xi_{1i}\nonumber\\
V_{a}(T,T_4,A)&=&T_p(2\pi\alpha')(k_{4a}+k_a)\nonumber\\
G_{ab}(A)&=&\frac{i \delta^{ab}}{(2\pi\alpha')^2T_p u}\\
V_b(A,T_2,T_3)&=&T_p(2\pi\alpha')(k_{2}-k_{3})_{b}\nonumber\\
G(T)&=&\frac{i}{(2\pi\alpha')T_p(u+r+w+1)}\labell{ver2}\eeqa
and $k$ is the off-shell tachyon's momentum. Replacing the above vertices to \reef{amp777xz} we would reproduce its double pole as well.

Note that  By direct calculations, the presence of some new couplings such as  $F^{(1)}\cdot F^{(2)}$ or  $D\phi^{i(1)}\cdot D\phi_{i(2)}$ has been confirmed in the world volume of D-brane-Anti D-brane systems \cite{Garousi:2007fk} and \cite{Hatefi:2016yhb}. Indeed making string calculation we could produce all massless and tachyon singularities of the amplitudes.

\vskip.2in

While WZ coupling $C_{p}\wedge DT \phi$ will not receive any higher derivative correction, and all  the kinetic terms are fixed, hence they do not  get any corrections either. Thus  all other  tachyon singularities give us clues about structures of all order higher derivative corrections to various  couplings and in this paper we could consistently fix their coefficients for good.

\vskip.2in

 Note that all these couplings are found in the limit  $p_ap^a\rightarrow 1/4$, thus we cannot compare  these couplings with BSFT couplings. However, tachyon's potential remains the same as  in BSFT ($V({T})=e^{\pi\alpha'm^2{ T}^2}$ ~\cite{Kutasov:2000aq}) which is \beqa
V( T^iT^i)&=&1+\pi\alpha'm^2{ T^iT^i}+
\frac{1}{2}(\pi\alpha'm^2{ T^iT^i})^2+\cdots
\non\eeqa  where 
$m^2=-1/(2\alpha')$ is tachyon's mass. Tachyon condensation is going to be carried out at $T\rightarrow \infty$, therefore its potential will be sent to zero. 
\vskip.1in

We think these corrections play crucial role in determining singularities of the higher point functions of string theories. Veneziano amplitude  \cite{Veneziano:1968yb} was generalised in  \cite{Hatefi:2017ags}, we hope to be able to address supersymmetric generalisation of the D-brane-anti-D-brane system by directly carrying out fermionic amplitudes \cite{EHsanPetr2018}. We also hope to have progress on the generalisation of  the non-supersymmetric DBI and WZ effective actions in near future.

\section*{Acknowledgements}

Some parts of the paper were done  at Vienna University of Technology as well as Queen Mary University of London. EH would like to thank K. Narain, L. Alvarez-Gaume for discussions and supports, also thanks to Mathematical Institute at Charles university for the hospitality. He is also grateful to IHES, CERN, UC Berkeley and Caltech for the warm hospitality. We thank A. Sagnotti, J. Polchinski, B. Jurco, O. Lechtenfeld, N.Arkani-Hamed, P. Horava, P. Sulkowski, G.Veneziano, J. Schwarz and W. Siegel for their insights and fruitful discussions. This work was supported  by an ERC Starting Grant no. 335739
"Quantum fields and knot homologies", funded by the European Research Council under the
European Union's 7th Framework Programme. EH was also supported in part by Scuola Normale Superiore and by INFN.

  \end{document}